\documentclass{article}

\usepackage{microtype}
\usepackage{graphicx}
\usepackage{subcaption}
\usepackage{booktabs}
\usepackage{array}

\usepackage{hyperref}

\usepackage[preprint]{icml2026}

\usepackage{amsmath}
\usepackage{amssymb}
\usepackage{mathtools}
\usepackage{amsthm}
\usepackage{physics}

\usepackage[capitalize,noabbrev]{cleveref}

\theoremstyle{plain}

\theoremstyle{definition}

\theoremstyle{remark}

\usepackage[textsize=tiny]{todonotes}

\icmltitlerunning{Accelerating Feedback-based Algorithms
for Quantum Optimization Using Gradient Descent}

\begin{document}

\twocolumn[
  \icmltitle{Accelerating Feedback-based Algorithms
for 
\\ Quantum Optimization Using Gradient Descent}

  \begin{icmlauthorlist}
    \icmlauthor{Masih Mozakka}{ind}
    \icmlauthor{Mohsen Heidari}{ind}
  \end{icmlauthorlist}

  \icmlaffiliation{ind}{Luddy School of Informatics, Computing, and Engineering, Indiana University Bloomington, Bloomington, IN, United States}
  \icmlcorrespondingauthor{Masih Mozakka}{mmozakka@iu.edu}

  \icmlkeywords{Quantum Optimization, QAOA, Quantum Lyapunov Control, Gradient Descent}

  \vskip 0.3in
]

\printAffiliationsAndNotice{}  

\begin{abstract}
Feedback-based methods have gained significant attention as an alternative training paradigm for the Quantum Approximate Optimization Algorithm (QAOA) in solving combinatorial optimization problems such as MAX-CUT. In particular, Quantum Lyapunov Control (QLC) employs feedback-driven control laws that guarantee monotonic non-decreasing objective values, can substantially reduce the training overhead of QAOA, and mitigate barren plateaus. However, these methods might require long control sequences, leading to sub-optimal convergence rates. 

In this work, we propose a hybrid method that incorporates per-layer gradient estimation to accelerate the convergence of QLC while preserving its low training overhead and stability guarantees. By leveraging layer-wise gradient information, the proposed approach selects near-optimal control parameters, resulting in significantly faster convergence and improved robustness. We validate the effectiveness of the method through extensive numerical experiments across a range of problem instances and optimization settings.

\end{abstract}

\section{Introduction}

The Quantum Approximate Optimization Algorithm (QAOA), introduced in \cite{farhi2014quantumapproximateoptimizationalgorithm}, is a hybrid classical-quantum framework for solving combinatorial optimization problems using parameterized quantum circuits. First the problem is encoded as a Hamiltonian so that its ground state gives the solution to the problem. Then, QAOA applies sequence of alternating quantum operations, whose variational parameters are optimized via a classical outer loop to minimize a target cost observable. The algorithm can be viewed as a discretized, variational counterpart of adiabatic quantum evolution, with increasing circuit depth enhancing expressive power. An approximate solution to the original optimization problem is obtained by measuring the optimized QAOA circuit in the computational basis.

QAOA has been extensively studied for a range of NP-hard optimization problems, including Minimum Vertex Cover and Maximum Independent Set, with applications spanning biology, finance, communications, healthcare, and machine learning \cite{sharma2025comparativestudyquantumoptimization, angara2025scoopquantumcomputingframeworkconstrained}. For a comprehensive overview of see \cite{BLEKOS20241}.

A central challenge in QAOA is the substantial quantum and classical resources required to optimize its variational parameters due to measurement-heavy gradient estimation at each iteration.  Moreover, as the circuit depth increases, the QAOA optimization landscape becomes increasingly nonconvex and can exhibit barren plateaus with exponentially vanishing gradients which severely hinder trainability \cite{Larocca2022}.

Quantum control frameworks have emerged as promising alternatives to QAOA for mitigating training challenges associated with variational quantum algorithms, including high measurement costs and difficult optimization landscapes \cite{PhysRevA.47.4593, 10.1063/1.1559680, Doherty_2000, ZHANG20171,Magann_2022, Magann_2022_2,PhysRevResearch.7.013035}. A prominent example is the feedback-based algorithm for quantum optimization (FALQON), which iteratively appends quantum circuit layers whose parameters are determined directly from measurement outcomes on the current quantum state \cite{Magann_2022,Magann_2022_2}. More generally, Quantum Lyapunov Control (QLC) employs  feedback laws derived from measuring a Lyapunov function to deterministically set circuit parameters, guaranteeing monotonic improvement of the objective without requiring an explicit classical optimization loop \cite{1272601, Magann_2022_2}.

Feedback-based algorithms have been extensively studied across a wide range of applications, including finding ground states of quantum Hamiltonians \cite{larsen2023feedbackbasedquantumalgorithmsground}, solving quadratic  and combinatorial optimization tasks \cite{Magann_2022,Magann_2022_2,BLEKOS20241}, and other applications in atomic physics and molecular chemistry \cite{Cong_2013, Brif_2010}, 

Despite their advantages, feedback-based algorithms often require long and finely discretized control sequences, which can lead to slow convergence and deep circuits that are challenging to implement on near-term quantum hardware. For example, in the weighted MAX-CUT problem on 3-regular graphs, the circuit depth required by FALQON to achieve accuracy comparable to QAOA has been shown to be up to two orders of magnitude larger \cite{Magann_2022_2}. Although a recent work has introduced second-order feedback laws to improve FALQON's performance \cite{PhysRevResearch.7.013035}, the fundamental challenge of slow convergence and large circuit depth remains.

This work integrates gradient-based optimization with FALQON to accelerate convergence and reduce circuit depth. Specifically, we incorporate gradient descent (GD) with a small number of iterations at each circuit layer to optimize the control parameters. This hybrid strategy enables faster convergence and shorter circuits while preserving the stability advantages of feedback control and helps avoiding challenges commonly encountered in QAOA training, such as barren plateaus.

We provide comprehensive numerical experiments validating our findings across a variety of optimization problems including MAX-CUT, MAX-CLIQUE and MIN-CLIQUE. The experiments  demonstrate that our method substantially improves purely feedback-based methods (FALQON) both in finding the global minima and stability under parameter variations.  Moreover, we analyze the impact of the number of per-layer GD iterations and the choice of timestep on algorithmic performance, showing that the proposed method exhibits strong robustness and well-behaved control parameters across a broad range of settings.

The rest of this paper is organized as follows: Section \ref{Preliminaries} briefly introduces QAOA, QLC, and FALQON. Section \ref{Main results} presents our main approach for convergence acceleration. Section \ref{numerical simulation} provides the numerical results and lastly Section \ref{conclusion} concludes the paper and discusses some future directions.

\section{Preliminaries} \label{Preliminaries}

\subsection{QAOA}

QAOA works by alternating applications of a problem-specific cost Hamiltonian $H_p$ and a driver (mixing) Hamiltonian $H_d$, with the goal of concentrating probability on high-quality solutions.  The Hamiltonians evolutions characterize the following unitary operations:
\begin{equation*}
    U_p(\gamma) = e^{- i \gamma H_p},\, U_d(\beta)=e^{- i \beta H_d},
\end{equation*}
with tunable parameters $\gamma$ and $\beta$.

Typically $H_p$ is designed so that  its ground state encodes the solution to the optimization problem. Starting with an initial state $\ket{s}$,  $K$ alternating applications of $U_p$ and $U_d$ with parameters ${\boldsymbol{\gamma}} = (\gamma_0,\, \gamma_1,\, \dots,\, \gamma_K)$ and ${\boldsymbol{\beta}} = (\beta_0,\, \beta_1,\, \dots,\, \beta_K)$ results in the state 
\begin{equation} \label{base qaoa relation}
    \ket{\psi({\boldsymbol{\gamma}},\, {\boldsymbol{\beta}})} = U_d({{\beta}}_K) U_p({{\gamma}_p})\cdots  U_d({{\beta}}_1) U_p({{\gamma}_1}) \ket{s}.
\end{equation}

A classical computer is used to iteratively tune the variational parameters ${\boldsymbol{\gamma}}$ and ${\boldsymbol{\beta}}$ in order to minimize the objective function
$$
E({\boldsymbol{\gamma}}, {\boldsymbol{\beta}}) = \bra{\psi({\boldsymbol{\gamma}}, {\boldsymbol{\beta}})} H_p \ket{\psi({\boldsymbol{\gamma}}, {\boldsymbol{\beta}})}.
$$
This optimization is typically carried out using GD-based update rules within a classical outer loop. The performance of QAOA is commonly quantified by the approximation ratio
$E_{\mathrm{QAOA}} / E_{min}$, which compares the expected cost value obtained from the QAOA output state to the optimal value corresponding to the ground state of $H_p$.

\subsection{Quantum Lyapunov Control}

Lyapunov control is a framework for steering the dynamics of a system toward a desired target by manipulating control parameters according to a feedback law derived from a \emph{control Lyapunov function}. Consider a controlled dynamical system of the form
\begin{equation*}
\dot{x} = f(x,u),
\end{equation*}
where \(x \in \mathbb{R}^n\) denotes the system state, $\dot{x}$ typically refers to its time derivative $\dv{x}{t}$, and \(u \in \mathbb{R}^m\) is the control input. The objective is to drive the system to a desired equilibrium point \(x^*\) from any initial state.

A continuously differentiable function \(V : \mathbb{R}^n \to \mathbb{R}_{\ge 0}\) is called a \emph{control Lyapunov function} if it satisfies: (i) \(V(x)\) is positive definite with respect to \(x^*\), i.e., \(V(x) > 0\) for all \(x \neq x^*\) and \(V(x^*) = 0\); and (ii) for every \(x \neq x^*\), there exists a control input \(u\) such that \(V\) is non-increasing along the system trajectories,
\begin{equation*}
\dot{V}(x) = \nabla V(x) \cdot f(x,u) \leq 0.
\end{equation*}
This condition guarantees that, for each state \(x\), an appropriate choice of control input can be made to decrease the Lyapunov function, thereby driving the system asymptotically toward the equilibrium \(x^*\).

Consider a quantum system whose dynamics are governed by the Schr\"odinger equation with \(\hbar = 1\):
\begin{equation}\label{base falqon equation}
i \frac{d}{dt} \ket{\psi(t)} = \bigl(H_p + \beta(t) H_d\bigr)\ket{\psi(t)},
\end{equation}
where \(\ket{\psi(t)}\) denotes the system state, and \(H_p\) and \(H_d\) are the  problem (drift) and driver (control) Hamiltonians, respectively, and \(\beta(t)\) acts as a control function. 

Suppose the objective is to find the ground state of $H_p$  minimizing the energy function:
\begin{equation} \label{energy of problem}
    E_p(t) = \bra{\psi(t)} H_p \ket{\psi (t)}.
\end{equation}

To apply QLC, a classical approach is to consider $E_p(t)$ as a Lyapunov \emph{candidate} for monotonic objective improvement \cite{1272601,Magann_2022}.  Let $E_0=\lambda_{\min}(H_p)$ be the minimum eigenvalue and define the shifted function 
\[
V(\ket{\psi}) := \bra{\psi}H_p\ket{\psi}-E_0 \ge 0,
\]
which equals $0$ on the ground-state of $H_p$. Since $\dot V = \frac{d}{dt}E_p(t)$, a feedback law that enforces $\frac{d}{dt}E_p(t)\le 0$ guarantees monotonic decrease of the objective, obtaining QLC law. Note that convergence to the ground state generally requires additional invariance/spectral conditions: for instance, non-degenerate eigenvalues and eigenvalue gaps of $H_p$) \cite{Magann_2022}.

\begin{figure*}[h]
    \vspace{-5pt}
    \centering
    \includegraphics[width=0.8\textwidth]{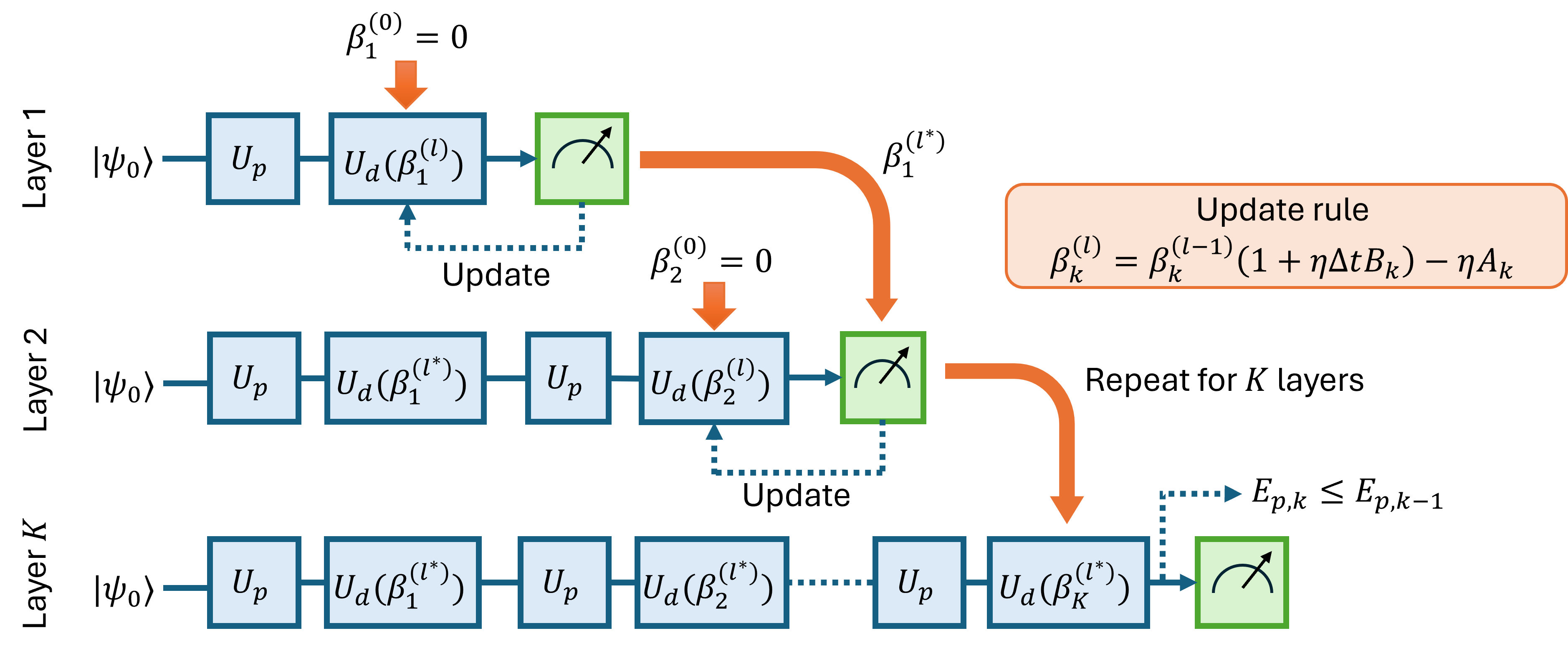}
    \vspace{-5pt}
    \caption{In our method, for each layer, GD is used to update $\beta_k^{(l)}$.  This process is repeated $K$ times to construct all the layers. Our method reduces to FALQON when GD is used for only one iteration (with $\eta=1$) in each layer.}
    \label{fig: gdqlc circuit}
\end{figure*}

In what follows, we present the feedback law explicitly.  Given the system evolving according to \eqref{base falqon equation}, one can show that the time derivative of $E_p(t)$ is 
\begin{align}
\frac{d}{dt} E_p
&= \expval{i[H_d, H_p]}{\psi(t)} \beta(t) \nonumber \\
&= A(t)\beta(t),
\label{falqon master}
\end{align}
where $A(t) :=\expval{i[H_d, H_p]}{\psi(t)}$.
Choosing the feedback control law $\beta(t) = -A(t)$ yields
\[
\frac{d}{dt} E_p = -A(t)^2 \leq 0,
\]
which satisfies the QLC condition and guarantees monotonic non-increasing evolution of the objective.
Building on this principle, Magann et al.~\cite{Magann_2022} introduced a time-discretized realization of this control strategy known as FALQON, where the continuous-time dynamics in \eqref{base falqon equation} are approximated by a sequence of discrete control steps of duration \(\Delta t\). Similar to QAOA, this method is implemented via alternating unitary evolutions under the control and problem Hamiltonians. Specifically, defining
\begin{equation}
U_{d,j} = e^{-i \Delta t \beta_j H_d}, \qquad
U_p = e^{-i \Delta t H_p},
\end{equation}
the state of the system after \(K\) steps, starting from an initial state \(\ket{\psi_0}\), is given by
\begin{equation*}
\ket{\psi_k}
= \prod_{k=0}^{K} U_{d, k}\, U_p \ket{\psi_0}.
\label{discrete falqon}
\end{equation*}

The resulting circuit is parameterized by the vector
\(\boldsymbol{\beta} = (\beta_0, \dots, \beta_K)\), which provides a piecewise-constant approximation to the continuous control field \(\beta(t)\). By iteratively appending layers and updating \(\beta_{k+1}\) using measurement-based estimates of $$A_{k} := A(2k\Delta t) = \bra{\psi_{k}} i [H_d,\, H_p] \ket{\psi_k},$$ FALQON constructs increasingly accurate approximations to the continuous-time Lyapunov control trajectory. Provided that $\Delta t$ is chosen sufficiently small, this discretization preserves the monotonic improvement guarantees of the underlying QLC framework.

\section{Main Results} \label{Main results}

Our main contribution is a hybrid optimization framework that combines QLC with per-layer GD to accelerate the convergence of feedback-based methods for combinatorial optimization problems. We refer to this approach as \emph{GD-QLC}. The central idea is to incorporate gradient information at each layer of the quantum circuit in order to refine the control parameters, thereby improving convergence rates while preserving the stability guarantees and low training overhead that characterize QLC.

Having $E_p$ as our Lyapunov function, an optimal control strategy over a finite time interval $[0, T]$ would ideally solve the following optimization problem:

\begin{equation*}
    \underset{\beta: \mathbb{R}^+\to \mathbb{R}}{\min}  E_p(T),
\end{equation*}
where the minimization is taken over all admissible control functions $\beta(t)$.

To this end,  at each layer $k$ (corresponding to time $t_k=k\Delta t$) in the discrete FALQON procedure, we propose to perform $L$ gradient descent updates on the control parameter $\beta_k$ before proceeding to the next layer (see Figure \ref{fig: gdqlc circuit}). In what follows, we derive the update rule.

Let $\dot{E}_{p,k}(\beta_k)=\pdv{E_p}{t}$ at time $t=(2k\Delta t)$ be the objective function of the GD at layer $k$. The goal is to minimize $\dot{E}_{p,k}(\beta_k)$ with respect to $\beta_k$. From \eqref{falqon master}, we know that $\dot{E}_{p,k}(\beta_k) = A_k(\beta_k)\beta_k$, where $A_k=A(2k\Delta t)$. Then the derivative of the objective with respect to $\beta$ is given by

\begin{align*}
    \dv{\dot{E}_{p,k}}{\beta}&= \dv{A_k(\beta)}{\beta} = \dv{A_k}{\beta}\beta+A_k.
\end{align*}

The derivative $\dv{A_k}{\beta}$ can be computed as follows:

\begin{align*}
  \dv{A_k}{\beta} &= \dv{}{\beta}\expval{i[H_d, H_p]}{\psi_k}\\
  &=\dv{}{\beta}i\expval{U^\dagger_pU_d^\dagger(\beta) [H_d, H_p]U_d(\beta)U_p}{\psi_{k-1}} \\
  &=-\Delta t \expval{[H_d, [H_d, H_p]]}{\psi_{k}},
\end{align*}
where we have used the fact that $\dv{U_d(\beta)}{\beta} = -i \Delta t H_d U_d(\beta)$ and the cyclic property of expectation values. 
Therefore, the derivative of the objective function at layer $k$ is given by
\begin{align*} 
    \dv{\dot{E}_{p,k}}{\beta} &= -\Delta t \expval{[H_d, [H_d, H_p]]}{\psi_{k}} \beta\\
    & + i\expval{[H_d, H_p]}{\psi_{k}}. 
\end{align*}
Let $B_k = \expval{[H_d, [H_d, H_p]]}{\psi_{k}}$, then the update rule for $\beta_k$ is given by
\begin{equation}  
    \beta_k^{(l+1)} = \beta_k^{(l)} (1+\eta(k,l) \Delta t B_k)- \eta(k,l) A_k, 
\end{equation}
where $\eta(k,l)$ is a learning rate that may depend on the layer index $k$ and the iteration index $l$. After performing $L$ such updates, the best $\beta_k^{(l^*)}$ is chosen and proceed to the next layer in the FALQON sequence. This process is summarized in Algorithm \ref{alg:special case of newfalqon}.
As for the choice of learning rate, we consider $\eta(k,l) = \frac{c}{\sqrt{l} \log k}$, where $c$ is a tunable constant. Our intuition is that as $k$ gets larger, we expect $\ket{\psi_k}$ to approach the desired solution implying that $\beta$ approaches zero.

It is important to note that GD-QLC performs updates independently at each layer, whereas QAOA repeatedly updates all circuit layers through a global iterative optimization loop. As a result, GD-QLC incurs substantially lower training overhead compared to standard QAOA. 

\subsection{Overhead comparison}
Setting $L = 1$ and $\eta(k,1) = 1$ recovers the FALQON method. More generally, the parameter $L$ controls the trade-off between convergence speed and measurement overhead, since each additional gradient step requires extra measurements to estimate the corresponding observables. Specifically, for each of the $K$ layers, $L$ gradient descent updates are performed in GD-QLC, each requiring the estimation of both $A_k$ and $B_k$. Thus, using $N_{\text{shot}}$ shots for each parameter estimation, the total number of measurements scales as $$O(K L N_{\text{shot}}).$$ This is  $L$ times higher than FALQON as it scales $O(K N_{\text{shot}}).$ We show in the next section that this additional overhead is justified by the significant reduction in the number of layers $K$ needed to achieve comparable performance.

The measurement overhead of training QAOA scales as $$O(N_{\mathrm{iter}}K N_{\mathrm{shots}}),$$ for  $N_{\mathrm{iter}}$ number of global gradient iterations. 
\begin{algorithm}[htb]
\caption{GD-QLC}
\label{alg:special case of newfalqon}
\textbf{Input}: $H_p, H_d$, $\ket{\psi_0}$, $\Delta t$, $K, L$
\begin{algorithmic}[1]

\STATE Define $U_p = e^{-i \Delta t H_p}$ and $U_d(\beta)=e^{-i \Delta t \beta H_d}$
\FOR{$k =1$ to $K$}
    \STATE Initialize $\beta^{(0)}_k = 0$
    \FOR{$l=1$ to $L$}
        \STATE Prepare the state $\ket{\psi_{k}} := U_d(\beta^{(l-1)}_k)U_p\ket{\psi_{k-1}}$
        \STATE Estimate $A_k = \expval{i[H_d,\, H_p]}{\psi_k}$ and  $B_k = \expval{[H_d, [H_d, H_p]]}{\psi_{k}}$.
        \STATE $\beta^{(l)}_k \leftarrow \beta^{(l-1)}_k (1+\eta(k,l) \Delta t B_k)- \eta(k,l) A_k$
    \ENDFOR
    \STATE Choose $\beta_k^{(l^*)}$ that minimizes $\dot{E}_p$ among $\beta_k^{l}$'s.
    \STATE Set  $\ket{\psi_k} = U_d(\beta^{(l^*)}_k)U_p\ket{\psi_{k-1}}$.
    \ENDFOR

\STATE Return $\beta_k^{(l^*)}$ and the state $\ket{\psi_K}$
\end{algorithmic}
\end{algorithm}

\begin{figure*}[ht]
  \begin{center}
    \centerline{\includegraphics[width=\textwidth]{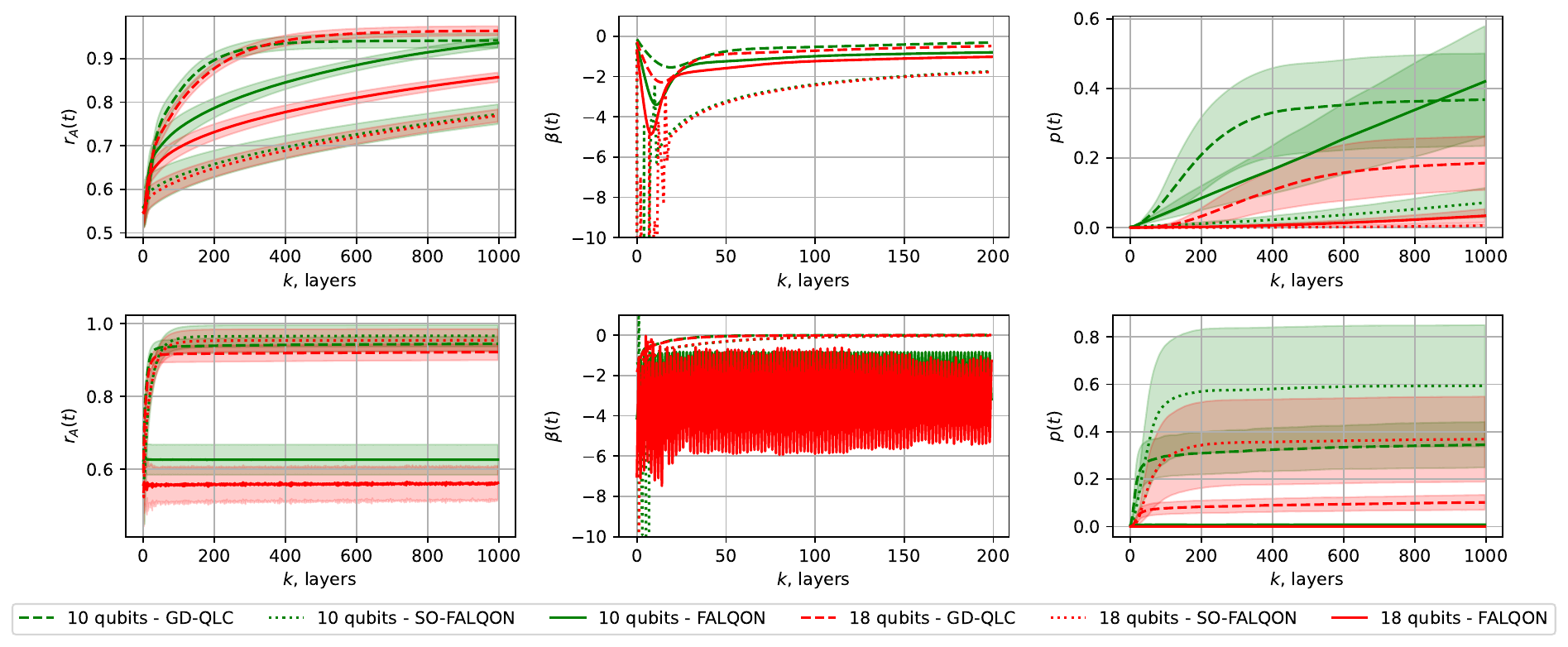}}
    \caption{
      Comparison of $R_A$ (left) and $p(t)$ (right) for solving weighted MAX-CUT using GD-QLC, FALQON, and SO-FALQON under two timestep settings: $\Delta t = 0.01$ (bottom row) and $\Delta t = 0.1$ (top row). GD-QLC exhibits robust performance across both regimes, outperforming FALQON and achieving performance comparable to SO-FALQON in the large-$\Delta t$ regime. In addition, GD-QLC displays significantly milder variations in the control parameters $\beta_k$ (middle column) compared to the other methods. For visualization purposes, the range of $\beta_k$ is clipped to the interval $[-10, 0]$, and results are shown for up to $200$ layers.
    }
    \label{fig: arai comparison}
  \end{center}
\end{figure*}

\section{Numerical Simulation} \label{numerical simulation}
To evaluate the performance of the proposed GD-QLC framework, we consider a set of well-studied graph-based combinatorial optimization problems: (1) MAX-CUT, (2) weighted MAX-CUT, (3) MAX-CLIQUE, and (4) MIN-COVER. These problems are natural benchmarks in the context of quantum optimization, as they admit QUBO formulations and can therefore be mapped to Ising Hamiltonians \cite{a12020034, Lucas_2014}. The corresponding problem Hamiltonians $H_p$ for each task are summarized in Table~\ref{graph ising table}.

For all problem instances, the initial state $\ket{\psi_0}$ is chosen to be the uniform superposition, and the driver Hamiltonian is fixed to
$$
H_d = \sum_i X_i,
$$
where $X_i$ is the Pauli-X on the $i$th register. 

\begin{table}[ht]
  \caption{Problem Hamiltonian for the studied graph problems}
  \vspace{-10pt}
  \label{graph ising table}
  \begin{center}
    \begin{small}
      \begin{sc}
        \begin{tabular}{m{2cm}l}
          \toprule
          Problem  &$H_p$ \\
          \midrule
          MAX-CUT   & $ \displaystyle \frac{1}{2}\sum_{(i,\,j)\in E(G)} (Z_i Z_j - I)$  \\
          weighted MAX-CUT & $ \displaystyle \frac{1}{2}\sum_{(i,\,j)\in E(G)} (w_{ij}Z_i Z_j - I)$  \\
          MAX-CLIQUE   & $ \displaystyle 3 \sum_{(i,\,j) \in E(\bar G)} Z_i Z_j - Z_i - Z_j + \sum_{i \in V(G)} Z_i$  \\
          MIN-COVER   & $ \displaystyle 3 \sum_{(i,\,j) \in E(G)} Z_i Z_j + Z_i + Z_j - \sum_{i \in V(G)} Z_i$  \\
          \bottomrule
        \end{tabular}
      \end{sc}
    \end{small}
  \end{center}
  \vspace{-10pt}
\end{table}

The proposed GD-QLC method is primarily compared against FALQON \cite{Magann_2022_2} and its second-order extension \cite{PhysRevResearch.7.013035}. We do not include standard QAOA baselines in our experimental comparisons, as FALQON and Arai’s variant already provide direct and comprehensive comparisons against QAOA and demonstrate improved performance in the relevant regimes \cite{Magann_2022_2, PhysRevResearch.7.013035}. 

In an idealized, noiseless setting with exact expectation values and exact gradients, QAOA can in principle outperform feedback-based methods. However, in practice, QAOA incurs a substantially higher training overhead due to the large number of measurement shots required for parameter optimization, which has been shown to significantly degrade its performance. In addition, as reported in \cite{Magann_2022_2, PhysRevResearch.7.013035}, the trainability of QAOA deteriorates with increasing circuit depth due to optimization landscape pathologies, including barren plateaus.

Given these practical limitations and the prior comparative studies, we focus our empirical comparisons on FALQON and Arai's method as the most informative baselines for the feedback-accelerated approach. All experiments were performed using classical simulations, and the source code is provided in the Supplementary Material.

\begin{figure*}[htb]
  \centering
    \includegraphics[scale=0.6]{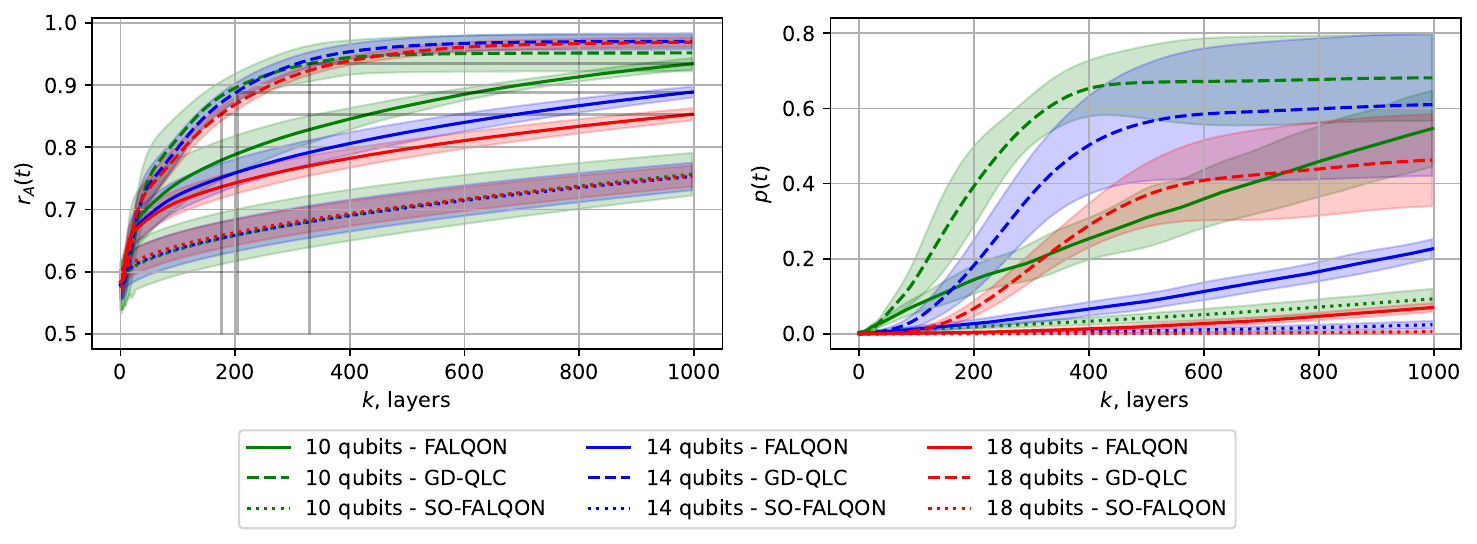}
  \caption{
    Performance comparison of GD-QLC, FALQON and SO-FALQON for solving MAX-CUT with $\Delta t = 0.01$. 
  }
  \label{fig:max-cut}
\end{figure*}

\begin{figure*}[htb]
  \centering
    \includegraphics[width=0.9\textwidth]{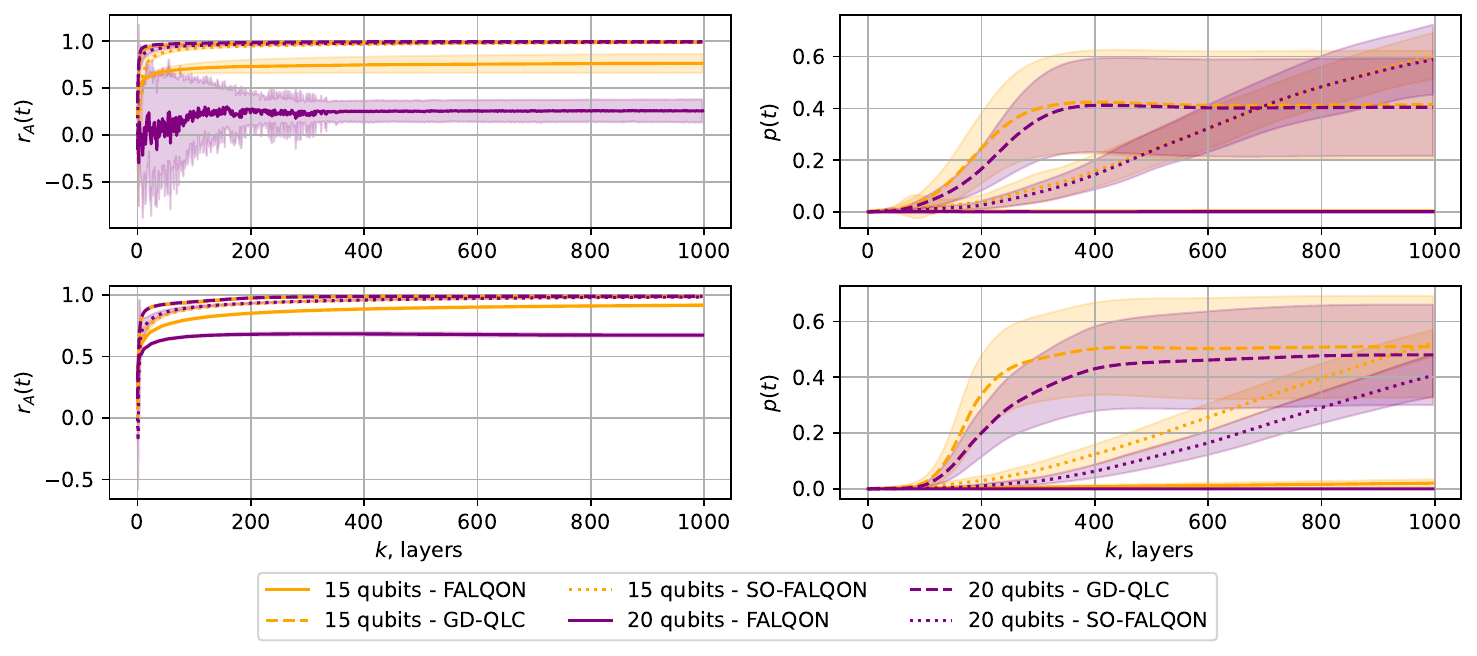}

  \caption{
    Performance comparison of GD-QLC against FALQON and SO-FALQON for MAX-CLIQUE (top row) and MIN-COVER (bottom row)  with $\Delta t = 0.005$.
  }
  \label{fig:max-clique-min-cover}
\end{figure*}

\subsection{Setup}
We consider 3-regular graphs for MAX-CUT, scale-free graphs for MIN-COVER, and random graphs for MAX-CLIQUE. Scale-free graphs are generated using the Barab\'asi--Albert model with parameter $m=3$, while random graphs are drawn from the Erd\H{o}s--R\'enyi model with edge probability $p=0.5$. For MAX-CUT, we include all non-isomorphic 3-regular graphs (see, e.g., \cite{regulargraphs}). 

For each qubit count and each method, we evaluate 40 randomly generated problem instances to compute averaged performance metrics, except in the case of 10-vertex 3-regular graphs, for which only 19 distinct graphs exist. For weighted MAX-CUT, edge weights were randomly chosen uniformly from the interval $[0,\,2]$, leading to an average expected weight of one.

We choose $\eta(k,l) = \frac{0.1}{\sqrt{l} \log k}$ as the learning rate, and $L=7$ for the number of iterations in GD-QLC. We choose $K=1000$ as the limit for the total number of layers for all the algorithms. Time steps $\Delta t$ and number of qubits vary for each experiment. 

The performance is evaluated as a function of the number of layers in terms of the approximation ratio and the success probability. The approximation ratio at step  $k$ is defined as
\begin{equation}
    r_A(k) = \frac{E_{p,k}}{E_{p, min}},
\end{equation}
where $E_{p, min}$ denotes the minimum eigenvalue of the problem Hamiltonian corresponding to the optimal solution. For problem formulations cast as maximization tasks, the objective is instead to identify the maximum eigenvalue, and the approximation ratio can be defined analogously.

\begin{figure*}[ht]
  \begin{center}
    \includegraphics[width=0.8\textwidth]{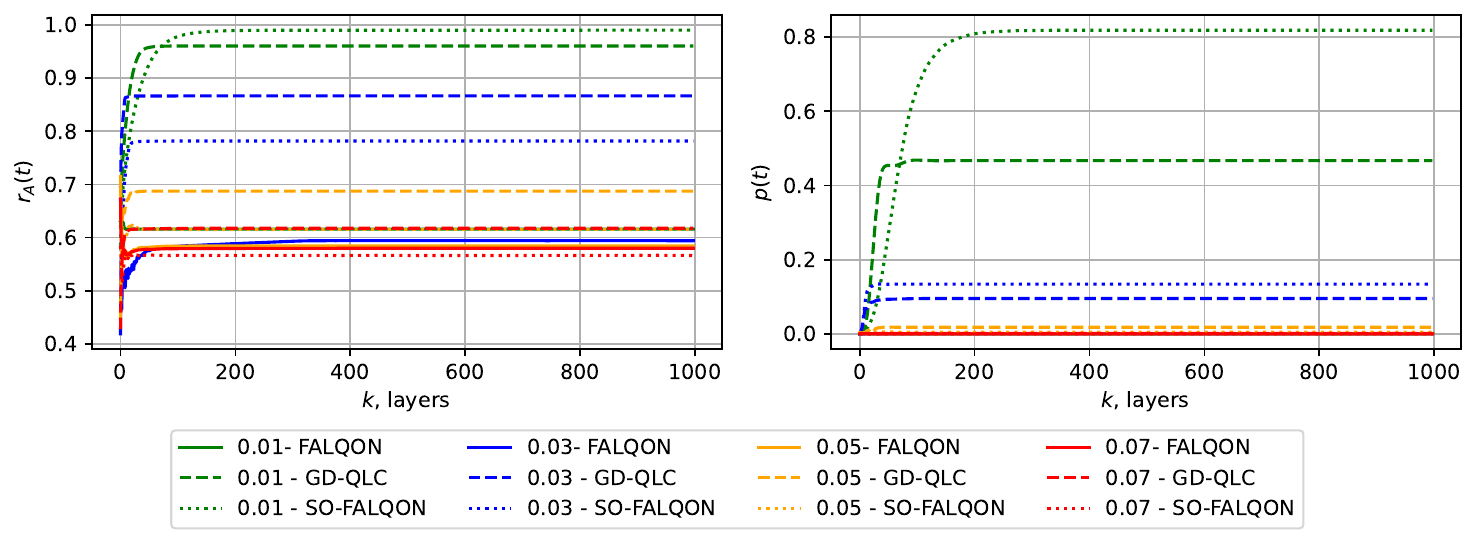}
    \caption{
      Results of solving weighted MAX-CUT using GD-QLC, FALQON and SO-FALQON with different timestep choices $\Delta t$. The results illustrate the robustness of GD-QLC for larger timesteps.
    }
    \label{fig: delta t comparison}
  \end{center}
\end{figure*}

The success probability $p(k)$ is defined as the probability of measuring a (potentially degenerate) ground state corresponding to a globally optimal solution of the original combinatorial optimization problem. It is given by
\begin{equation}
    p(k) = \sum_i \left|\braket{\psi_k}{q_i}\right|^2,
\end{equation}
where $\ket{q_i}$ denote the computational basis states that encode optimal solutions.

\subsection{Performance comparisons}
Figure~\ref{fig: arai comparison} compares GD-QLC with FALQON and SO-FALQON for solving weighted MAX-CUT at $\Delta t = 0.01$ and $\Delta t = 0.1$, which serve as representative benchmarks for small and large time steps, respectively. The left and right columns report the evolution of $r_A$ and $p(k)$ as functions of the number of layers, while the middle column shows the corresponding control parameters $\beta_k$.

Across both regimes, GD-QLC consistently outperforms FALQON. In the small-$\Delta t$ regime, GD-QLC also outperforms SO-FALQON, while achieving comparable performance in the large-$\Delta t$ regime.

SO-FALQON is expected to perform better at larger $\Delta t$ due to its second-order approximation. However, its feedback law selects extremely large values of $\beta_k$ in the early layers (see Figure~\ref{fig: arai comparison}), which poses challenges for practical implementations, as noted in the original work. 

In contrast, GD-QLC attains similar or better performance without exhibiting such instabilities. The control trajectories produced by GD-QLC remain well behaved across both timestep regimes, with substantially smoother and smaller-magnitude variations in $\beta_k$. Our numerical experiments show that for $\Delta t = 0.1$, FALQON and SO-FALQON can produce large spikes in $\beta_k$, exceeding $4000$ in magnitude, whereas GD-QLC avoids these extreme values while maintaining rapid convergence in both $r_A$ and $p(t)$. For visualization purposes, the range of $\beta_k$ is clipped in the figure.

Similar comparisons are presented for MAX-CUT, MAX-CLIQUE, and MIN-COVER in Figure~\ref{fig:max-cut} and Figure~\ref{fig:max-clique-min-cover} across a range of problem sizes (10--20 qubits). For the MAX-CUT problem, a timestep of $\Delta t = 0.01$ was used, while for MAX-CLIQUE and MIN-COVER a smaller timestep $\Delta t = 0.005$ was employed. This choice reflects the fact that typically $||H_{p,\, \text{MIN-COVER}}|| > ||H_{p,\, \text{MAX-CUT}}||$, and as a consequence, FALQON requires a smaller timestep to maintain stable performance. Similar considerations apply to the MAX-CLIQUE Hamiltonian.

Across all problem classes and sizes considered, the results demonstrate that GD-QLC consistently outperforms FALQON in terms of both the approximation ratio $r_A$ and the success probability $p(t)$. Larger problem sizes were not computationally tractable within classical simulation limits. The specific choice of qubit counts varies across problems and is constrained by the structure of the corresponding Hamiltonians and available computational resources.

\begin{figure*}[ht]
  \begin{center}
        \includegraphics[width=\textwidth]{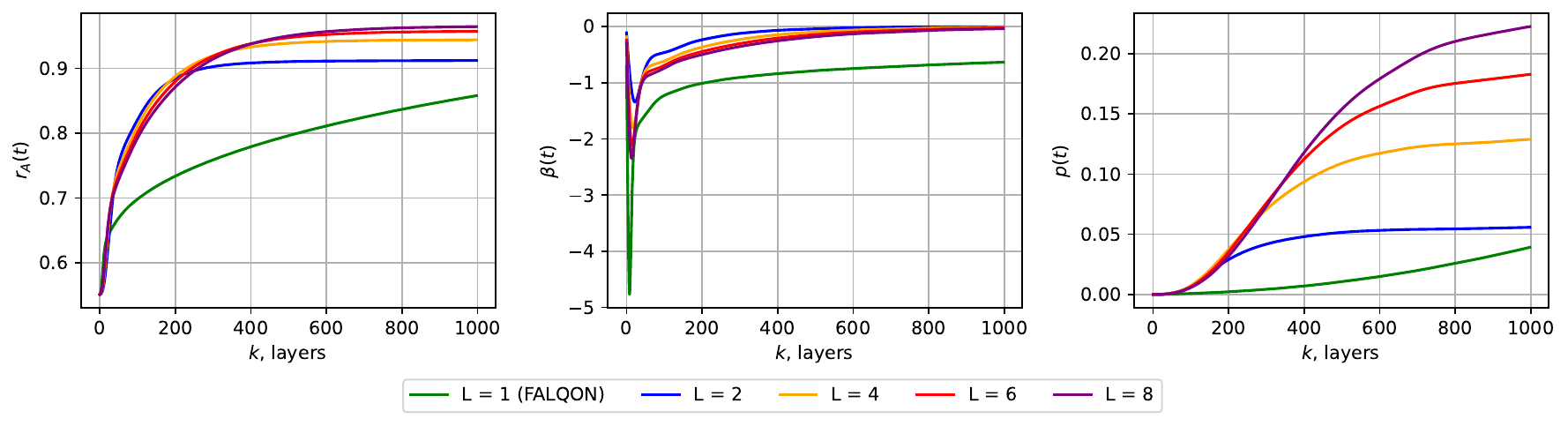}
    \caption{
      Results of solving weighted MAX-CUT using various choices of $L$ and $\Delta t = 0.01$. $L = 1$ is equivalent to FALQON as the update rule reduces to the FALQON case.
    }
    \label{fig: L comparison}
  \end{center}
  \vspace{-5pt}
\end{figure*}

\subsection{Robustness against  $\Delta t$}
Motivated by practical implementations in which fine-tuning the timestep may not be feasible, we further investigate the sensitivity of algorithmic performance to the choice of $\Delta t$ (Figure~\ref{fig: delta t comparison}). As discussed in \citep{Magann_2022}, achieving optimal performance with FALQON typically requires careful tuning of $\Delta t$ for each problem instance. When $\Delta t$ is too small, convergence becomes slow due to insufficient state evolution per layer. Conversely, when $\Delta t$ is too large, the linear approximations underlying the matrix exponentials break down, resulting in degraded performance for FALQON \cite{PhysRevResearch.7.013035}. SO-FALQON was proposed in part to mitigate this limitation through higher-order feedback laws.

We numerically demonstrate that the gradient-based updates in GD-QLC substantially improve robustness with respect to variations in $\Delta t$. Specifically, we compare $r_A$ and $p(t)$ for solving weighted MAX-CUT on 18 qubits across multiple timestep values $\Delta t = 0.01, 0.03, 0.05, 0.07$. Across all tested values of $\Delta t$, GD-QLC consistently reaches its optimal $r_A$ significantly faster than the other methods, typically within 500 layers. As expected, very small values of $\Delta t$ lead to slower convergence for all methods, since each layer induces only a weak evolution of the quantum state; nevertheless, GD-QLC maintains a clear advantage in convergence speed across the entire range of timesteps considered.

\subsection{Effect of $L$, the number of GD iterations}
Figure~\ref{fig: L comparison} illustrates the effect of the number of per-layer GD iterations $L$ on the performance of GD-QLC. We focus on the weighted MAX-CUT problem with $\Delta t = 0.01$, although similar behavior is expected for other problem instances. The left and right panels report the evolution of $r_A$ and $p(t)$, respectively, while the middle panel shows the corresponding control parameters $\beta_k$.

As $L$ increases, the performance of GD-QLC improves and converges toward a plateau, indicating diminishing returns beyond a moderate number of gradient steps. This behavior suggests the existence of a practical sweet spot for $L$ that balances improved convergence against the additional computational overhead introduced by per-layer GD updates. Consistent with the typically sharp initial improvements observed in GD, relatively small values of $L$ already yield substantial gains over the $L = 1$ case (i.e., FALQON), making the added overhead well justified. In addition, larger values of $L$ lead to significantly higher quality of solutions ($p(t)$), even when improvements in $r_A$ begin to saturate. 

Finally, the middle panel shows that larger values of $L$ also help regulate the control amplitudes, mitigating sharp transients in $\beta_k$ observed for small $L$. Taken together, these results indicate that per-layer GD not only accelerates convergence but also improves control stability and solution quality, with performance gains that saturate asymptotically as $L$ increases.

\section{Conclusion and Future Work} \label{conclusion}

In this work, we introduced GD-QLC, a hybrid optimization framework that integrates per-layer GD updates into feedback-based QLC methods to accelerate convergence and reduce circuit depth for combinatorial optimization problems. By combining the stability and low training overhead of QLC with local gradient information, our approach addresses a central limitation of existing feedback-based algorithms, namely their slow convergence and reliance on long control sequences.
 
Through extensive numerical simulations on MAX-CUT, weighted MAX-CUT, MAX-CLIQUE, and MIN-COVER, we demonstrated that GD-QLC consistently outperforms FALQON across a range of problem sizes in terms of both the approximation ratio $r_A$ and the success probability $p(t)$. We further showed that GD-QLC exhibits strong robustness to variations in the timestep $\Delta t$, achieving rapid convergence without the need for careful tuning or higher-order feedback laws. This robustness is  relevant for near-term quantum hardware, where fine-grained control  and long circuit depths may be impractical.

\noindent\textbf{Future work.}
Several directions merit further investigation. From a theoretical perspective, it would be valuable to establish formal convergence-rate guarantees for GD-QLC and to characterize the conditions under which layer-wise gradient refinement preserves Lyapunov monotonicity.  Extending the framework to second‑order optimization techniques, such as natural gradient descent, or incorporating multiple control Hamiltonians and adaptive driver selection may enhance both expressivity and convergence speed. Finally, experimental validation on quantum hardware and applications to broader classes of optimization and control problems remain important directions for future work.

\section*{Acknowledgments}
This work was partially supported by NSF Grant CCF-2211423. This research was supported in part by Lilly Endowment, Inc., through its support for the Indiana University Pervasive Technology Institute.

\bibliography{main_arxiv.bbl}
\bibliographystyle{icml2026}

\newpage
\appendix
\onecolumn

\end{document}